\documentclass[aps,
twocolumn,
superscriptaddress
]{revtex4}%
\usepackage{amsfonts}
\usepackage{amsmath}
\usepackage{amssymb}
\usepackage{graphicx}%
\usepackage[dvipsnames]{xcolor}
\begin{document}

\title{Effect of 
density of states 
peculiarities 
on Hund's metal behavior 
}

\author{A. S. Belozerov}
\affiliation{M. N. Mikheev Institute of Metal Physics, Russian Academy of Sciences, 620137 Yekaterinburg, Russia}
\affiliation{Ural Federal University, 620002 Yekaterinburg, Russia}

\author{A. A. Katanin}
\affiliation{M. N. Mikheev Institute of Metal Physics, Russian Academy of Sciences, 620137 Yekaterinburg, Russia}

\author{V. I. Anisimov}
\affiliation{M. N. Mikheev Institute of Metal Physics, Russian Academy of Sciences, 620137 Yekaterinburg, Russia}
\affiliation{Ural Federal University, 620002 Yekaterinburg, Russia}

\begin{abstract}
\vspace{0.0cm}
We investigate a possibility of Hund's metal behavior in the Hubbard model with asymmetric density of states having peak(s). Specifically, we consider the degenerate two-band model and compare its results to the five-band model with realistic density of states of iron and nickel, showing that the obtained results are more general, provided that the hybridization between states of different symmetry is sufficiently small. We find that quasiparticle damping and the formation of local magnetic moments due to Hund's exchange interaction are enhanced by both, the density of states asymmetry, which yields stronger correlated electron or hole excitations, and the larger density of states at the Fermi level, increasing the number of virtual electron-hole excitations. For realistic densities of states these two factors are often interrelated because the Fermi level is attracted towards peaks of the density of states.
We discuss the implication of the obtained results to various substances and compounds, such as transition metals, iron pnictides, and cuprates. 
\end{abstract}
\maketitle

\section{Introduction}
Some components of Coulomb interaction, in particular, Hund's exchange, play special role in 
multiorbital systems, since they 
induce electronic correlations,
which may drive an unusual behavior of the electronic degrees of freedom. Hund's metals were defined \cite{HM} as metals with large quasiparticle mass, 
caused by Hund's exchange interaction, cf. Refs. \cite{Medici,Medici1,Pruschke,SF}. 
Apart from enhancement of quasiparticle damping (and the possibility of non-quasiparticle behavior), Hund's interaction also yields in the vicinity of half filling the so-called 'spin freezing' (i.e. local moment) behavior, as was shown in Ref.~\cite{SF} for semicircular density of states (DOS), and investigated for a number of real materials \cite{Licht2001,OurAlpha0,OurAlpha,Belozerov2013,Pourovskii,Sangiovanni,OurGamma,Toschi,OurPnic}. 
This behavior can be considered as one of the distinctive features of Hund's metals and it is characterized by the temperature-independent  local spin correlation function $K(\tau)=\langle S_i(\tau) S_i(0) \rangle$
for the imaginary time $\tau$ not too close to $0$ or $\beta=1/T$ ($i$ is the site index, $T$ is the temperature). The $\tau$-dependence of the correlation function $K(\tau)$ in this regime is rather weak, which then yields (approximate) fulfillment of the Curie law for the static local susceptibility $\chi_{\rm loc}=\int_0^\beta K(\tau) d \tau\propto 1/T$,  corresponding to a local moment formation. At the transition to the spin freezing (i.e. local moment) phase, the non-Fermi-liquid ($\omega^{1/2}$) behavior of electronic self-energy has been obtained \cite{SF}, while inside the spin freezing phase the quasiparticle damping is essentially enhanced. 

In realistic substances and compounds only electrons belonging to part of the bands may show 
Hund's metal
behavior, which is reminiscent of the orbital-selective Mott transitions, introduced first to explain unusual properties of Ca$_{2-x}$Sr$_x$RuO$_4$ \cite{Anisimov} and studied actively within the dynamical mean-field theory \cite{OSMT1,OSMT2,OSMT3,OSMT4,OSMT5,OSMT6,OSMT7}. In this respect, the ``orbital selective'' Hund's metals can be considered as being in proximity to the orbital-selective Mott transitions. The prominent example is $\alpha$-iron, which has a non-quasiparticle form of $e_g$ states \cite{OurAlpha0}, while the $t_{2g}$ states are more quasiparticle-like, although they also show some deviations from the Fermi liquid behavior~\cite{OurAlpha}. 

While for conventional orbital-selective Mott transitions, the widths of the bands and/or their fillings are the decisive factors in which bands 
undergo the transition, the widths and fillings of different bands in ``orbital-selective'' Hund's metals are close to each other, and the states, exhibiting Hund's metal behavior (including the local moment formation), are determined to a large extent by the  
profile of the partial density of states. Indeed, the $e_{g}$ electrons in $\alpha$-iron, for which the above described features of Hund's metal behavior
are especially pronounced,
have the Fermi level almost at the top of the peak of the density of states; 
in $\gamma$-iron \cite{OurGamma} and pnictides \cite{OurPnic,Pnic1,Pnic2,Pnic3,Pnic4,Pnic5}
the peak of the density of states is somewhat shifted with respect to the Fermi level, which is accompanied by weaker (in comparison with $\alpha$-iron)  quasiparticle damping and only partially formed local moments \cite{OurGamma,OurPnic,Toschi}. 
Therefore, the questions can be posed as to which factors (apart from the filling) are decisive for that,
which states (if any) show Hund's metal behavior in these substances, and whether the proximity of the Fermi level to the peak of the density of states is a necessary/sufficient condition for that. 

In the present paper we show that the quasiparticle damping and formation of local moments are in fact enhanced by the asymmetry of the density of states (which often shifts the Fermi level
towards the maximum of the density of states), as well as by the larger 
value of the density of states at the Fermi level. 
In the presence of the asymmetry, the behavior of the self-energy and spin 
correlation function above- and below half filling is drastically different; the Hund's metal behavior is enhanced for the Fermi level being on the same side, as the maximum of the density of states. 
We also find that Hund's exchange interaction is necessary to make these factors active,
similarly to a previous observation for the symmetric density of states \cite{SF}.

The plan of the paper is the following. In Sec. II we consider the results for the self-energies and local susceptibilities of the two-band model. In Sec. III we consider the results of the five-band model and implications of the obtained results to real substances and compounds. Finally, in Sec. IV we present conclusions.

\section{Two-band model}
To investigate the effects of the peculiarities
of the density of states on Hund's metal behavior, we perform dynamical mean-field theory (DMFT) calculations for the degenerate two-band Hubbard model (we have verified that the three-band model yields similar results)
\begin{eqnarray} \label{full_ham}
H &=& \sum_{ijm\sigma} t_{ij}c^{+}_{im\sigma}c_{jm\sigma} +U \sum_{im} n_{im\uparrow} n_{im\downarrow}  \\
&+& \hspace{-0.75em} \sum_{i,m > m^\prime,\sigma 
}
\! \left[ U' n_{im\sigma} n_{im^\prime\overline{\sigma}}
 +  (U'-I) n_{im\sigma} n_{im^\prime\sigma} \right], \nonumber
\end{eqnarray}
where $c_{im\sigma}$ ($c^+_{im\sigma}$) are the electron destruction (creation) operators with spin~$\sigma$ (${=\uparrow,\downarrow}$) at site $i$ and orbital~$m=1,2$; $n_{im\sigma}$ is the number operator of electrons, $U$ is the intraorbital Coulomb interaction,
$I$ is the Hund's coupling, and ${U'=U-2I}$.
To obtain the correlation strength similar to that in Hund's metals such as pure iron and iron pnictides, we consider the interaction values 
${U=1.5 D}$ and ${I=U/4}$, where $D$ is half of the bandwidth.
%
%
The impurity problem was solved by the hybridization expansion continuous-time quantum Monte Carlo method~\cite{CT-QMC}. 



\begin{figure}[t]
\includegraphics[clip=true,trim=-0.97cm 0cm 0cm 0cm, width=0.44\textwidth]{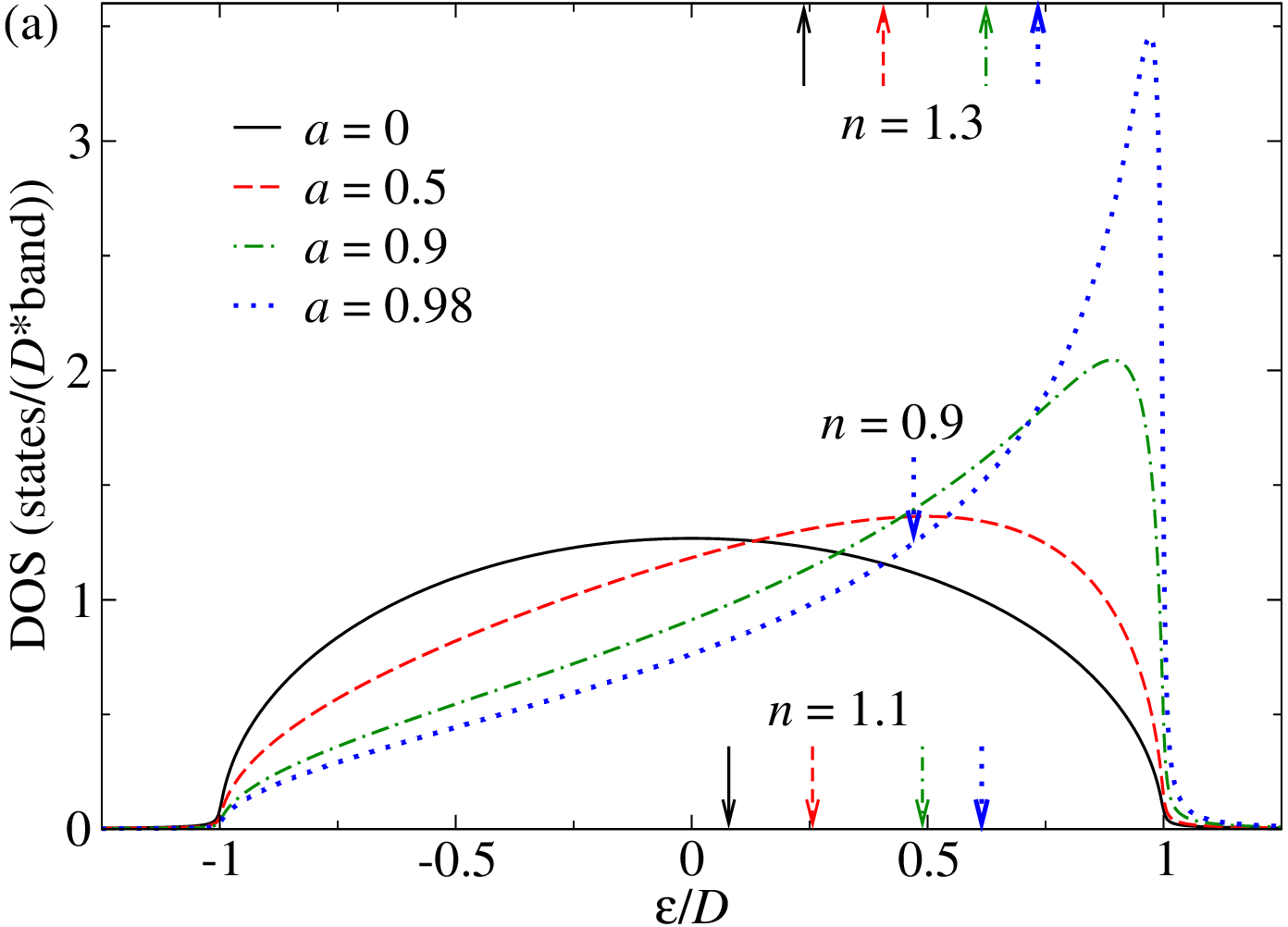}

\includegraphics[clip=true,width=0.441\textwidth]{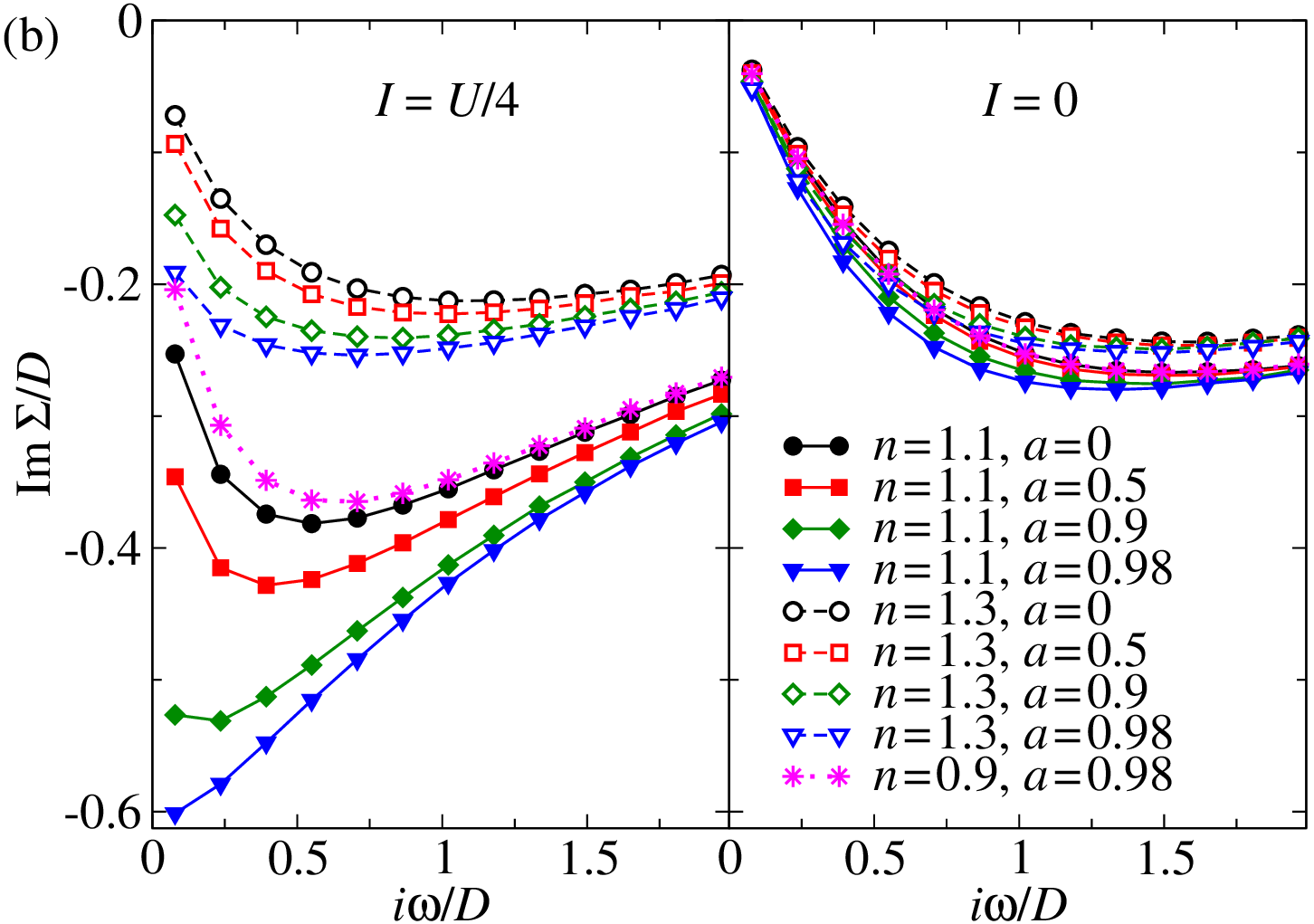}
\caption{(Color online)
\label{model_with_peak}
(a) Density of states $\rho_a(\varepsilon)$ for different values of the asymmetry parameter $a$. The positions of the respective Fermi levels are shown by 
arrows in the middle of the figure (for filling ${n=0.9}$), and at the lower (for ${n=1.1}$) and upper (for ${n=1.3}$) axes (the value of $a$ increases from left to right 
arrows).
(b) The respective imaginary parts of the self-energies on the Matsubara frequency axis with (left panel) and without (right panel) Hund's exchange. 
}
\end{figure}

Let us first consider the results for the 
DOS $\rho_a(\varepsilon)=c\,\sqrt{D^2-\varepsilon^2}/(D-a\varepsilon)$ with $c=(1+\sqrt{1-a^2})/(\pi D)$, suggested in Ref.~\cite{Wahle} and leading to a peak in the density of states for $a$ close to one (see Fig.~\ref{model_with_peak}a). Note that in this and the following calculations, we enforce the paramagnetic state by assuming spin- and site independent self-energy, since we are interested in the formation of local moments (see below). In Fig.~\ref{model_with_peak}b we present the imaginary parts of self-energies obtained with band fillings ${n=1.1}$ and ${n=1.3}$ (here and below the band fillings are indicated per band) at the inverse temperature $\beta=1/T=40 D^{-1}$. One can see that for both fillings increasing asymmetry parameter $a$
yields a larger absolute value of the imaginary part of the self-energy; for ${n=1.1}$ and strongest asymmetry ${a=0.98}$ the absolute value of the self-energy is even increasing with decreasing Matsubara frequency, corresponding to the non-quasiparticle behavior, similar to that, observed for $e_g$ states in $\alpha$ iron~\cite{OurAlpha0}. We note that the effective bandwidth of the density of states $\rho_a(\varepsilon)$, characterized by the second central moment (standard deviation) $\sigma^2=(1/2)\int d\varepsilon (\varepsilon-\overline{\varepsilon})^2 \rho(\varepsilon)$ with respect to the mean value of the energy $\overline{\varepsilon}=(1/2)\int d\varepsilon \varepsilon \rho(\varepsilon)$, is $\sigma=D/2$ and does not depend on $a$. As we argue below, the observed strengthening of local correlations with increasing $a$
can be explained by two effects: increasing asymmetry itself, which can be characterized by the skewness of the DOS $\alpha=1/(2 \sigma^3)\int d\varepsilon (\varepsilon-\overline{\varepsilon})^3 \rho(\varepsilon)$, changing from $\alpha=0$ at $a=0$ to $ \alpha=-0.82$ at $a=0.98$, and increase of the density of states at the Fermi level.
The former effect yields narrowing of the holes band width $W_h$, defined as a distance from the Fermi level to the upper edge of the band 
(in the case when the major weight of the density of states is below the center of the band, the bandwidth for electrons $W_e$, corresponding to the distance from the Fermi level to the lower edge, is narrowed instead), while the latter effect increases the number of virtual particle-hole excitations. 

The asymmetric form of the density of states with the peak also yields strong difference of the frequency dependence of the self-energy above and below half filling. To illustrate this, we also present in Fig.~\ref{model_with_peak}b the results for ${n=0.9}$ and strong asymmetry, ${a=0.98}$. One can see that the absolute value of the imaginary part of the self-energy in this case is much smaller, 
than for ${n=1.1}$, 
${a=0.98}$ and has the quasiparticle-like imaginary frequency dependence.
Without Hund's exchange interaction we find that all of the above discussed peculiarities of the self-energy disappear (see Fig.~\ref{model_with_peak}b) and the self-energy depends rather weakly on $n$ and $a$. Therefore, similarly to the previous study of the symmetric density of states, Hund's exchange interaction represents a driving force of the obtained anomalies of the self-energy, which is due to
the decrease of the critical value of Coulomb interaction for the metal-insulator transition by Hund's exchange near half filling
\cite{Medici1,Medici,Medici1a,Medici2,Medici3}.

\begin{figure}[t]
\includegraphics[clip=true,trim=-0.18cm 0cm 0cm 0cm,width=0.441\textwidth]{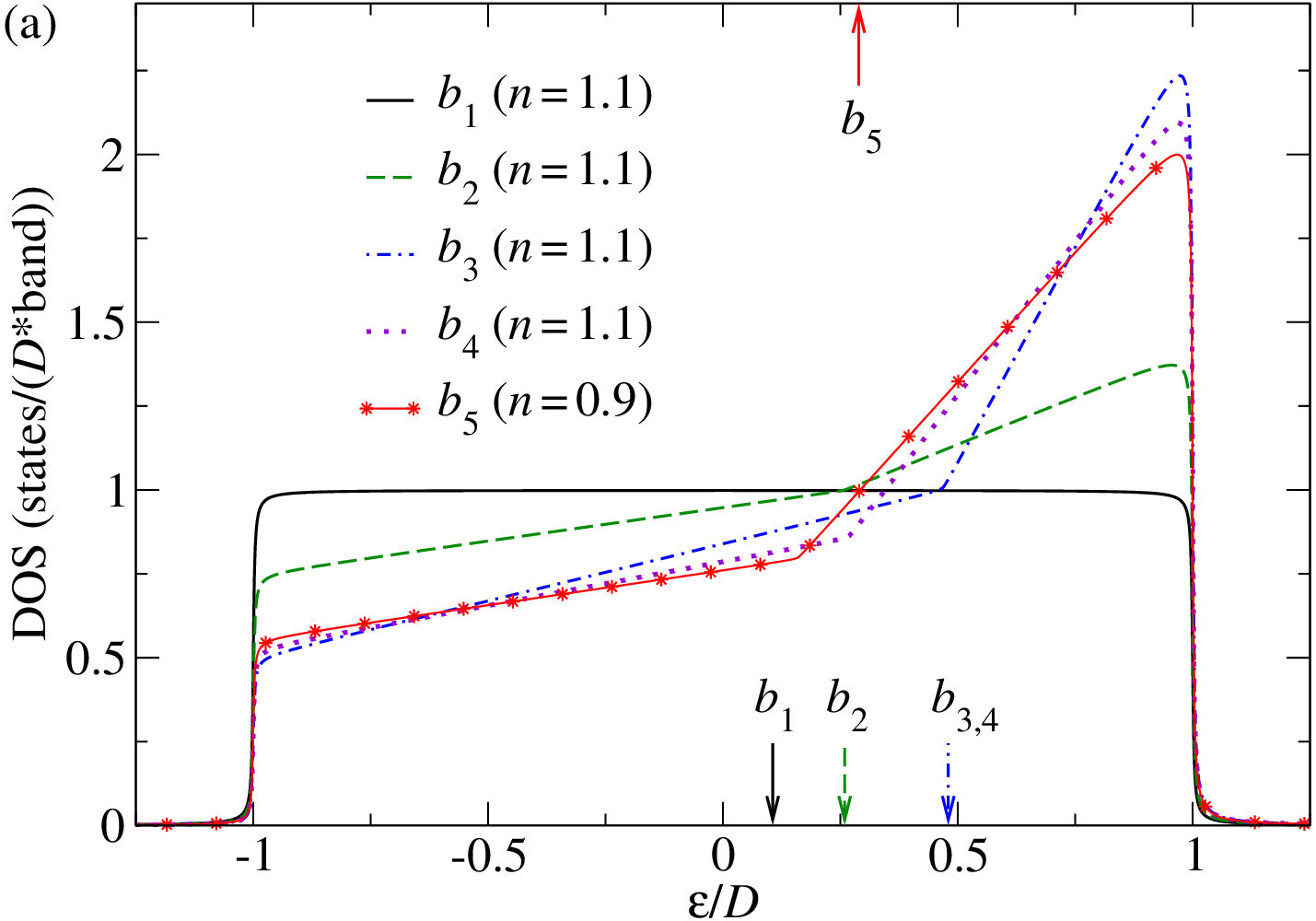}
\includegraphics[clip=true,width=0.44\textwidth]{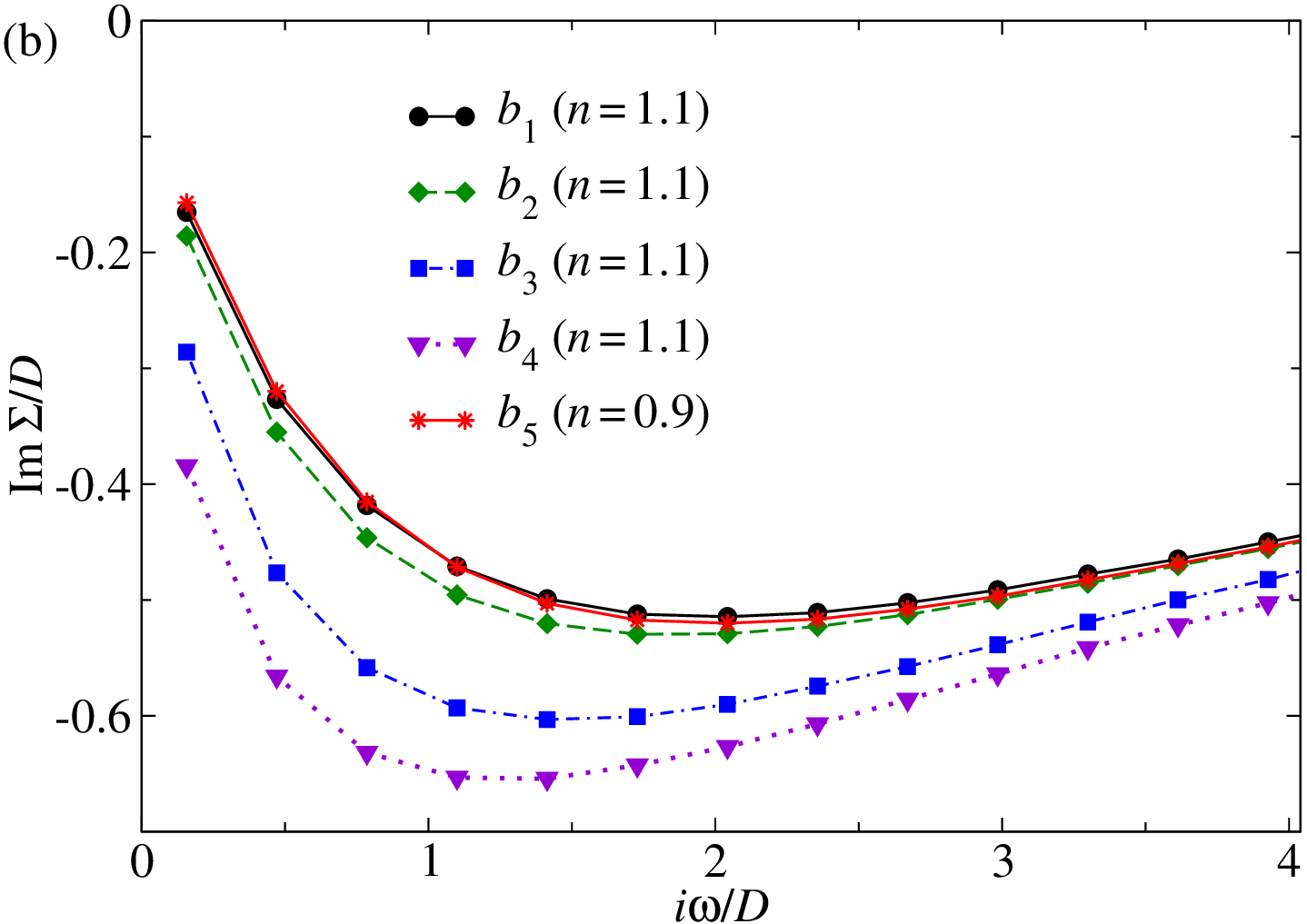}
\caption{
(Color online)
\label{model_with_asym1}
The same as in Fig. 1 for the 
density of states $\rho_b$, allowing us to disentangle the effects of asymmetry and the value $\rho(E_\textrm{F})$. The arrows indicate the positions of the corresponding Fermi levels.
}
\end{figure}

As it is mentioned above, the considered 
density of states 
at the Fermi level $\rho_a(E_\textrm{F})$ increases with increasing asymmetry. 
To disentangle the effects of the asymmetry and increasing of the density of states, 
we consider one more model density of states $\rho_b(\varepsilon)$, consisting of two linear energy dependences at $\varepsilon<E_\textrm{F}$ and $\varepsilon>E_\textrm{F}$. 
We fix the value of the density of states at the Fermi level, but change the asymmetry (see dependences $\rho_{b_{1,2,3}}$ on Fig.~\ref{model_with_asym1}a); we have verified that this yields an almost unchanged effective bandwidth $\sigma \approx 0.58D$. One can see that increasing the asymmetry in this case
yields a qualitatively similar
enhancement of $|{\rm Im}\Sigma|$
as in 
the above discussed results for the density of states $\rho_a(\varepsilon)$. 
Somewhat weaker enhancement in this case can be explained by the relatively weak asymmetry ($\alpha= -0.22$ and $\alpha= -0.50$ for $\rho_{b_2}$ and $\rho_{b_3}$, respectively) and smaller value of the density of states at the Fermi level. 
Increasing the value of the density of states at the Fermi level (see $\rho_{b_4}(\varepsilon)$ on Fig.~\ref{model_with_asym1}a), but keeping the standard deviation $\sigma$ 
and skewness $\alpha$ the same as for the density of states  $\rho_{b_3}(\varepsilon)$ (the position of the Fermi level also almost does not change), yields further enhancement of $|{\rm Im}\Sigma|$. 
Therefore, both factors, i.e., the asymmetry of the density of states and the value of the density of states at the Fermi level
enhance quasiparticle damping.
At the same time, increasing filling 
yields decreasing the absolute value of the imaginary part of the self-energy (see $n=1.1$ and $n=1.3$ results on Fig.~\ref{model_with_peak}), despite the fact that the Fermi level approaches the maximum (peak) of the density of states, since the strength of correlations decreases away from half filling. This shows that a shift of the filling away from half filling plays a more important role in this case than an increase of the density of states. 
Changing the filling to $n=0.9$ and keeping the density of states at the Fermi level, the second moment $\sigma$ and skewness $\alpha$ the same as for the density of states  $\rho_{b_3}(\varepsilon)$ (see the density of states $\rho_{b_5}(\varepsilon)$), we find almost the same quasiparticle damping, as for the fully symmetric band with filling $n=1.1$, which also agrees with the results for the density of states $\rho_a(\varepsilon)$ shown in Fig. 1. This suggests that the asymmetry enhances quasiparticle damping only for electron (hole) excitations 
for the position of the Fermi level on the same side from half filling as the maximum of DOS and sufficiently far from the center of the band, which supports the suggested mechanism of narrowing of the bandwidth for hole $W_h$ (or electron $W_e$) excitations.

\begin{figure}[t]
\includegraphics[clip=true,width=0.45\textwidth]{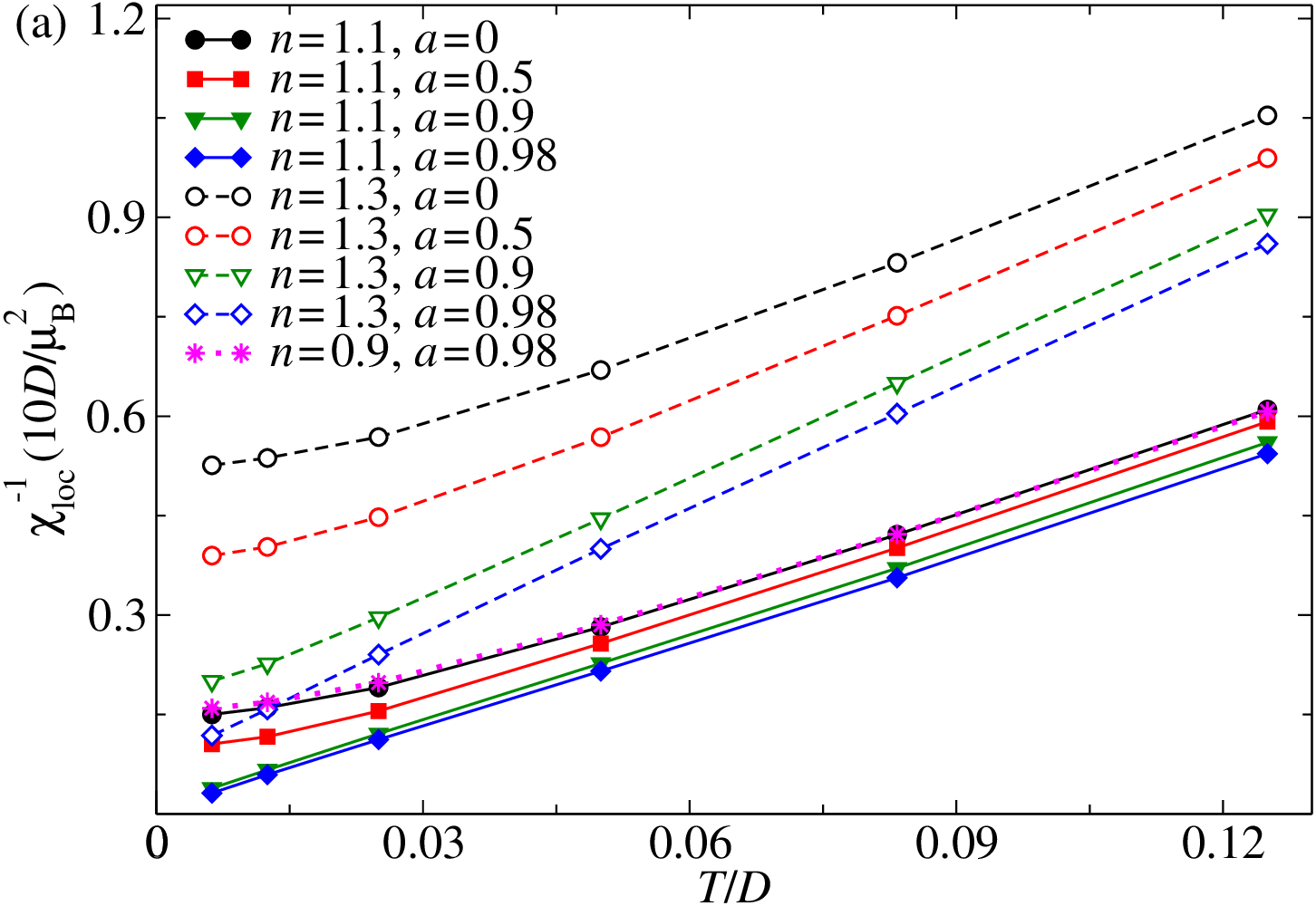}\vspace{0.3cm}
\includegraphics[clip=true,trim=-0.50cm 0cm 0cm 0cm,width=0.456\textwidth]{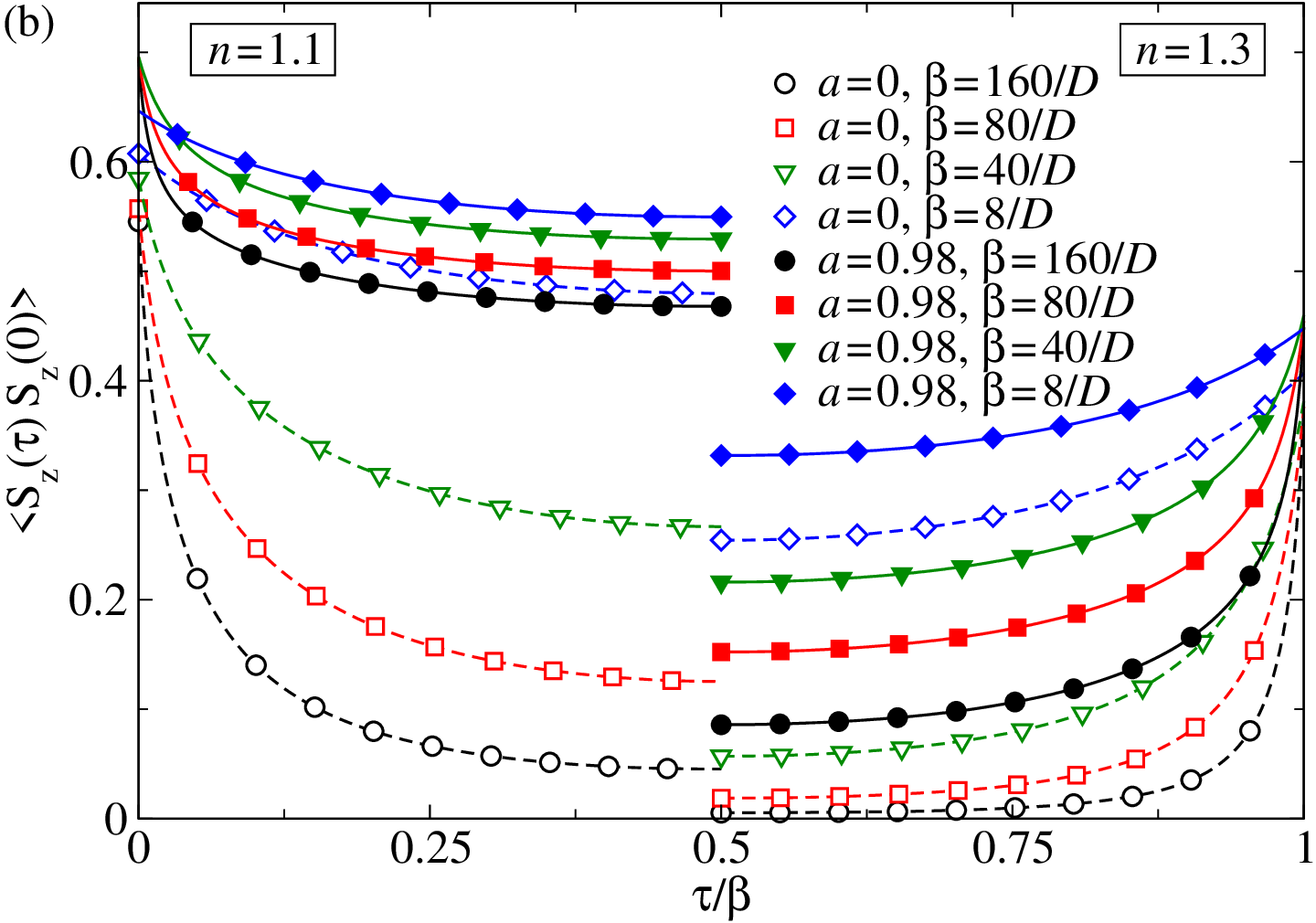}
\includegraphics[clip=true,trim=-0.50cm 0cm 0cm 0cm,width=0.456\textwidth]{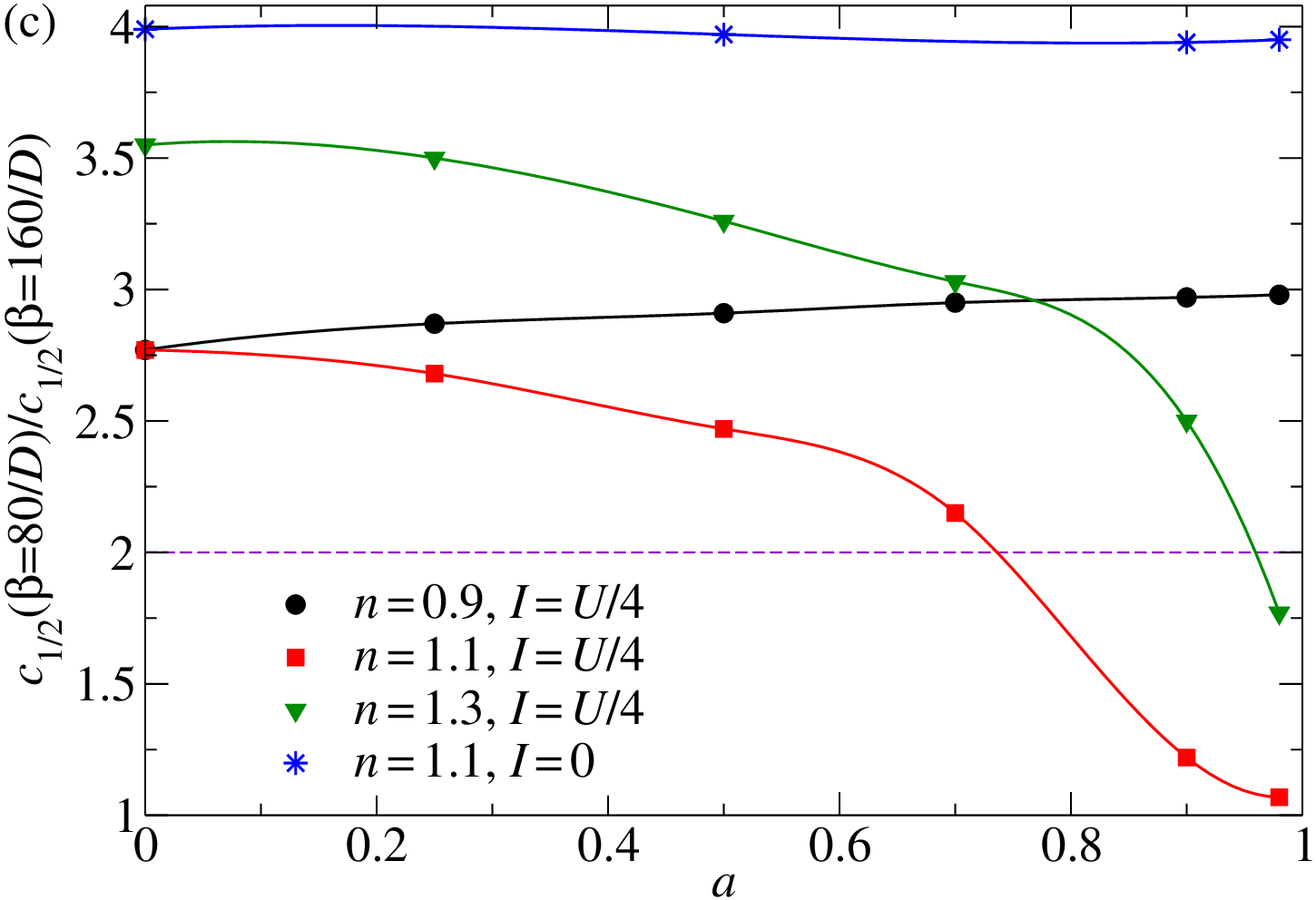}
\caption{(Color online)
\label{Fig:chi_loc}
(a) Temperature dependence of the inverse local susceptibility for the parameters considered in Fig.~\ref{model_with_peak} with ${I=U/4}$.
(b) Local spin-spin correlation functions in the imaginary-time domain. (c) The dependence of the ratio of spin correlators $c_{1/2}(\beta)=K(\beta/2)$ at $\beta=80D^{-1}$ and $\beta=160D^{-1}$  on the asymmetry parameter $a$. The solid lines were obtained by spline interpolation. The horizontal dashed line represents the criterion for the spin-freezing transition~\cite{SF}.
}
\end{figure}

To show the effect of the obtained self-energies on the formation of local moments, in Fig.~\ref{Fig:chi_loc}a we show the temperature dependence of the inverse local susceptibility in the presence of Hund's exchange for the density of states $\rho_a$ and model parameters, shown in Fig.~\ref{model_with_peak}. One can see that above half filling $\chi_{\rm loc}^{-1}(T)$ becomes linear for strong asymmetry of the density of states, ${a=0.98}$, when $|{\rm Im} \Sigma (0)|$ has its maximal value, while $\chi_{\rm loc}^{-1}(T)$ shows a crossover between Pauli-like and linear behavior for smaller $a$.  The size of the local moment, determined by the slope of $\chi_{\rm loc}^{-1}$, decreases going away from half filling, and almost does not depend on the asymmetry of the density of states. 
The Kondo temperature $T_K$,   corresponding to the temperature scale, below which local moments are screened by itinerant electrons (note that for the considered paramagnetic state within DMFT the Kondo effect does not compete with magnetism)
can be determined from the offset of the inverse susceptibility, $\chi^{-1}_{\rm loc} \propto T+2T_K$, similarly to the single local moment 
case \cite{Wilson,Note}, cf. Ref. \cite{Sangiovanni}. The obtained $T_K$ moderately decreases with increasing asymmetry of the density of states, but strongly increases going away from half filling.

Studying the $\tau$-dependence of the local spin-spin correlation function (see Fig.~\ref{Fig:chi_loc}b), we also observe the features of spin freezing (i.e. local moment formation) for those cases, when the inverse local susceptibility is linear in temperature: the $\tau$-dependence of the correlation function becomes weak and away from $\tau=0,\beta$ it is weakly temperature dependent, cf. Refs. \cite{Licht2001,SF,OurAlpha0,Toschi,Sangiovanni}. From the requirement of two times difference (which corresponds to the criterion, suggested in Ref.~\cite{SF}) of $K(\beta/2)$ at ${\beta=80/D}$ and ${\beta=160/D}$ we find the transition to the spin freezing (local moment) phase at $a\approx 0.74$ for ${n=1.1}$ and ${a\approx 0.96}$ for ${n=1.3}$ (see Fig.~\ref{Fig:chi_loc}c).


The local moments (when they exist) interact via RKKY-type of exchange, cf. Refs. \cite{OurAlpha,Alpha_supercell}, that may induce some type of magnetic order, which formation we do not study here. The feedback of this exchange to the formation of the local moments is however expected to be small, since the latter is provided by Hund's exchange on the much larger energy scale, than the non-local magnetic interactions. This allows us to neglect non-local effects in the present study.

Finally, we note that in the considered model we did not introduce the hybridization between different bands, but weak hybridization is not expected to change the obtained results.
In the presence of several types of states of different symmetry (e.g. $e_g$ and $t_{2g}$ for the cubic symmetry), the obtained results can again be qualitatively applied to these states separately, if the hybridization between the states of different symmetry is sufficiently small, and they are not strongly mixed by the Coulomb interaction (see below).  

\section{Five-band model and implications to various substances and compounds}

\begin{figure}[t]
\includegraphics[clip=true,trim=-0.41cm 0cm 0cm -0.06cm, width=0.436\textwidth]{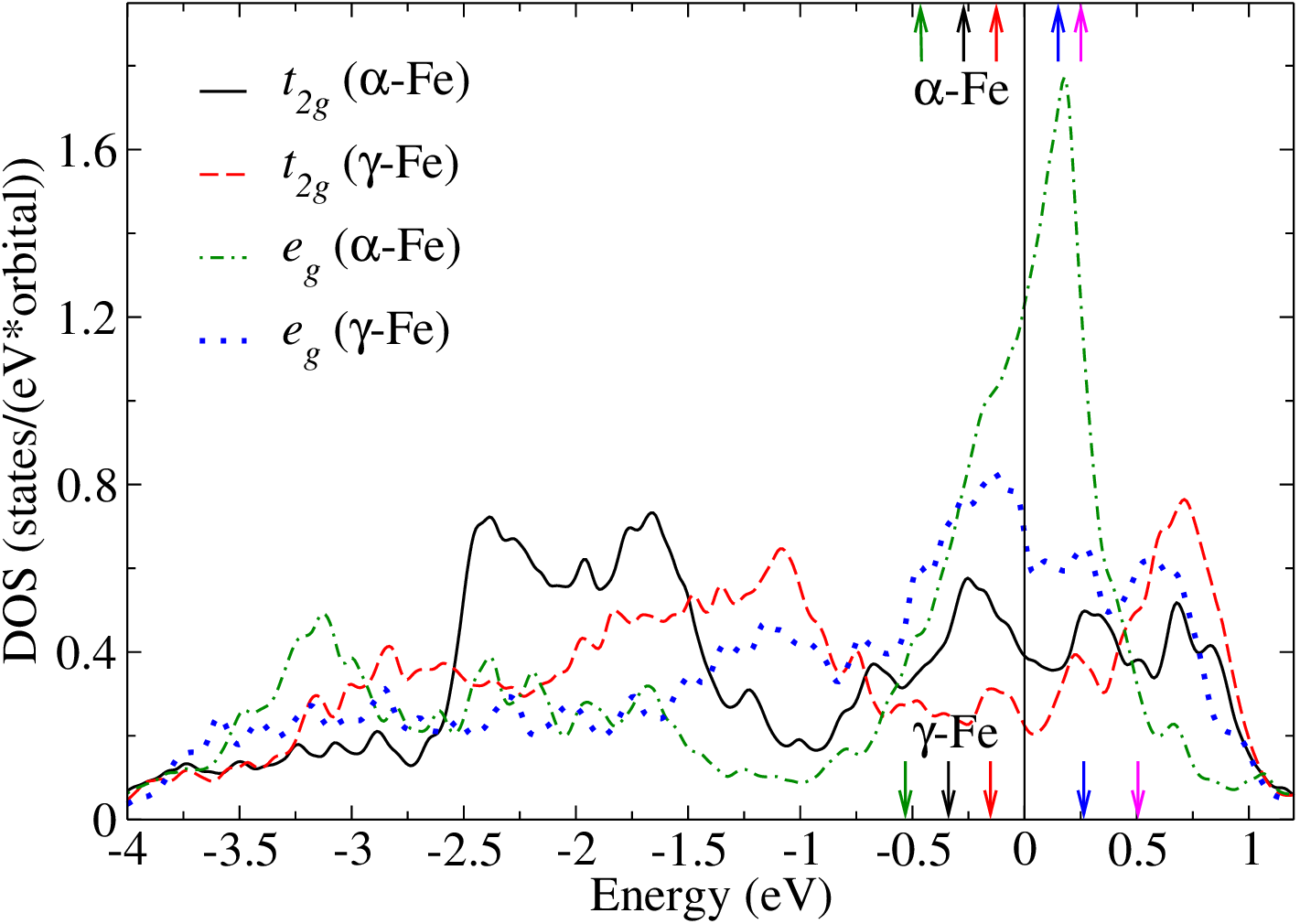}\vspace{0.16cm}
\caption{(Color online)
\label{fig:Fe_dos}
Density of $t_{2g}$ and $e_g$ states of $\alpha$- and $\gamma$-Fe, obtained by GGA. The arrows at the top (bottom) axis correspond to positions of the chemical potential for various fillings in $\alpha$-Fe ($\gamma$-Fe).
The vertical line corresponds to the position of the Fermi level in pure iron.  
}
\end{figure}

Let us discuss the effect of the 
density of states peculiarities
for the substances and compounds, having partially formed local moments. Iron in $\alpha$ and $\gamma$ phases has almost the same filling of the $d$-states 6.78 and 6.76, respectively, close values of standard deviation $\sigma$ of $t_{2g}$ and $e_g$ states, but $\alpha$-iron has stronger asymmetry of the density of 
states and larger value of the density of states at the Fermi level (see Fig. \ref{fig:Fe_dos} and Table I), and therefore more pronounced local moment behavior. In both substances the quasiparticle damping and spin freezing are stronger for the $e_g$ than $t_{2g}$ states because of stronger asymmetry and larger density of states at the Fermi level, but also the slight difference of fillings (${n^\alpha_{e_g}=1.24}$, ${n^\alpha_{t_{2g}}=1.44}$,
${n^\gamma_{e_g}=1.28}$,  ${n^\gamma_{t_{2g}}=1.40}$),
such that the $e_g$ orbitals are closer to half filling. 
We note that in contrast to the two-band model, discussed above, in these and the following results we account for the finite hybridization between different states. As it is usual in the \textit{ab initio} calculations, for the states of the same symmetry we interpret the results in terms of the (partial) densities of states. The hybridization between states of different symmetry is sufficiently small for $\alpha$-iron and nickel and moderate for $\gamma$-iron (see the Appendix), and the consideration in terms of partial densities of states for $t_{2g}$ and $e_g$ states remains valid, especially for the two former substances. 

\begin{figure}[t]
\includegraphics[clip=true,width=0.44\textwidth]{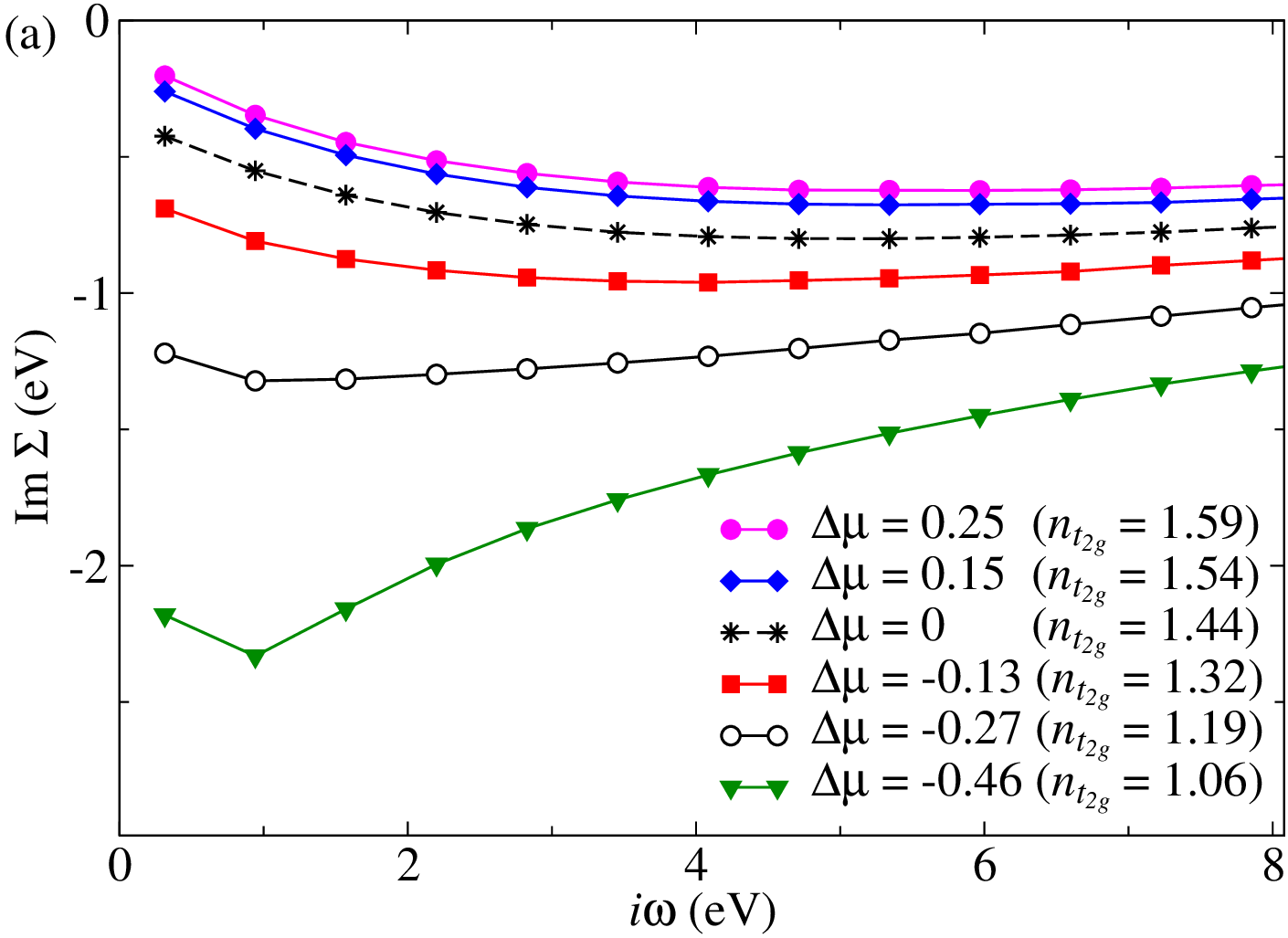}
\includegraphics[clip=true,width=0.44\textwidth]{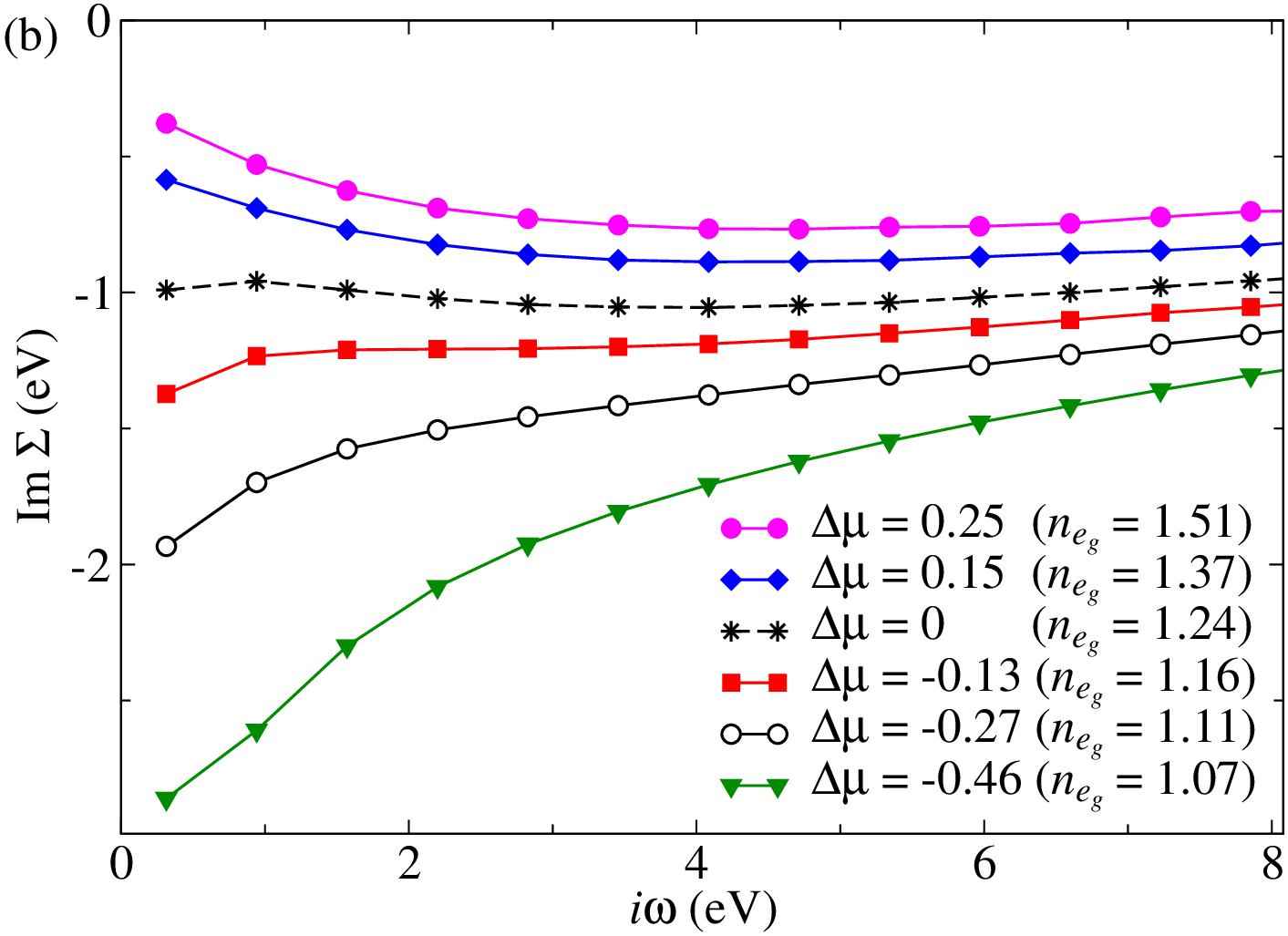}
\caption{(Color online) Self-energies of $t_{2g}$ (a) and $e_g$ (b) states of $\alpha$-iron for different fillings and $\beta=10$~eV$^{-1}$. 
\label{fig:sigma_bcc}
}
\end{figure}

\begin{table}[b]
\begin{ruledtabular}
\begin{tabular}{cccc}
Substance    &  States    & $\sigma$ (eV) & $\alpha$ \\
\hline
$\alpha$-Fe  &  $t_{2g}$  &    1.36       &     $-$0.012         \\ 
 &  $e_{g}$   &    1.43       &     $-$0.596         \\ 
$\gamma$-Fe  &  $t_{2g}$  &    1.40       &     $-$0.068         \\
 &  $e_{g}$   &    1.37       &     $-$0.545         \\ 
Ni           &  $t_{2g}$  &    1.26       &     $-$0.036         \\
         &  $e_{g}$   &    1.22       &     $-$0.479         \\ 
\end{tabular}
\end{ruledtabular}
\caption{Standard deviation (third column) and skewness coefficient (fourth column) of density of states for
$\alpha$-Fe, $\gamma$-Fe and Ni.}
\end{table}

\begin{figure}[t]
\includegraphics[clip=true,width=0.44\textwidth]{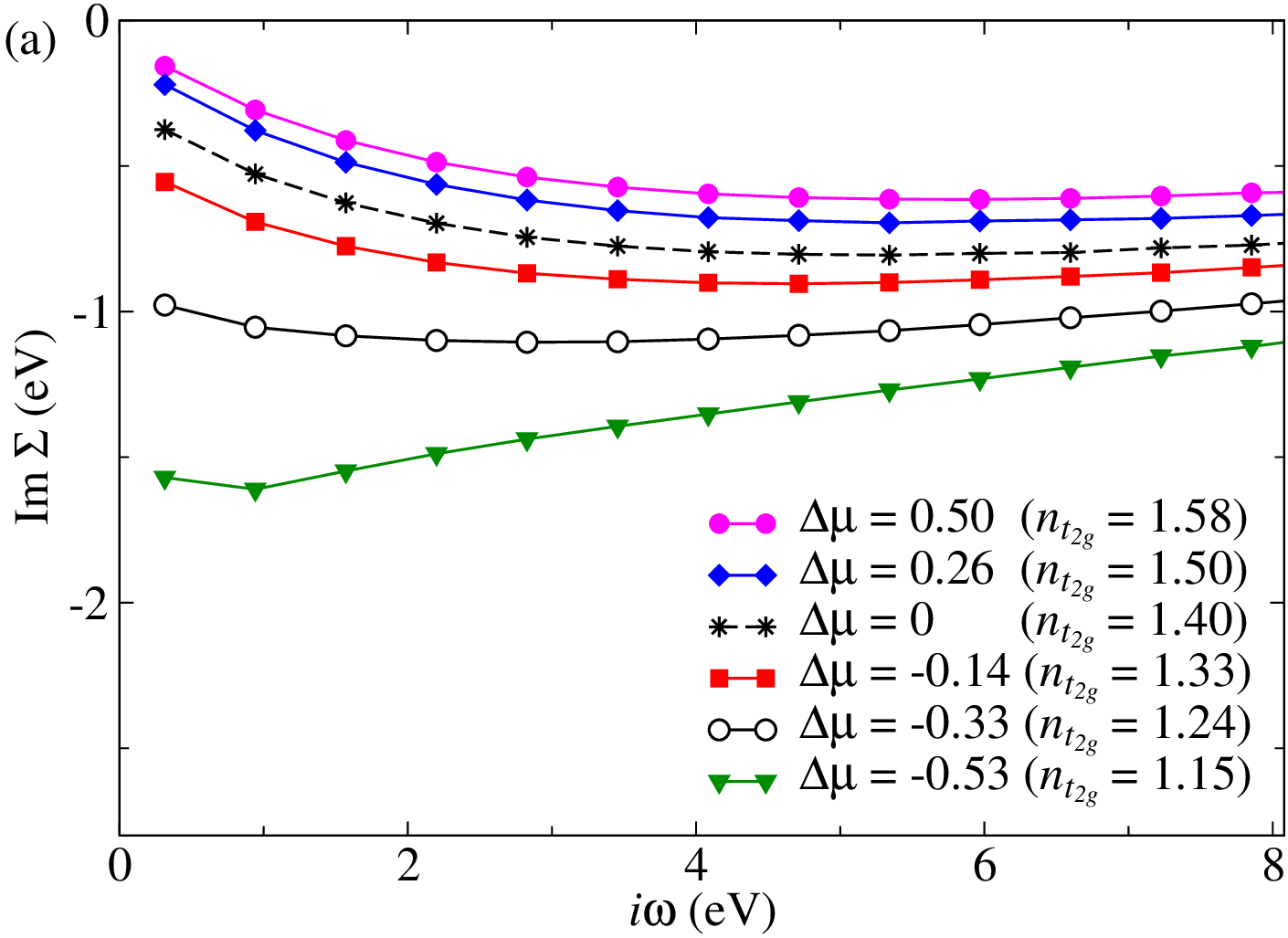}
\includegraphics[clip=true,width=0.435\textwidth]{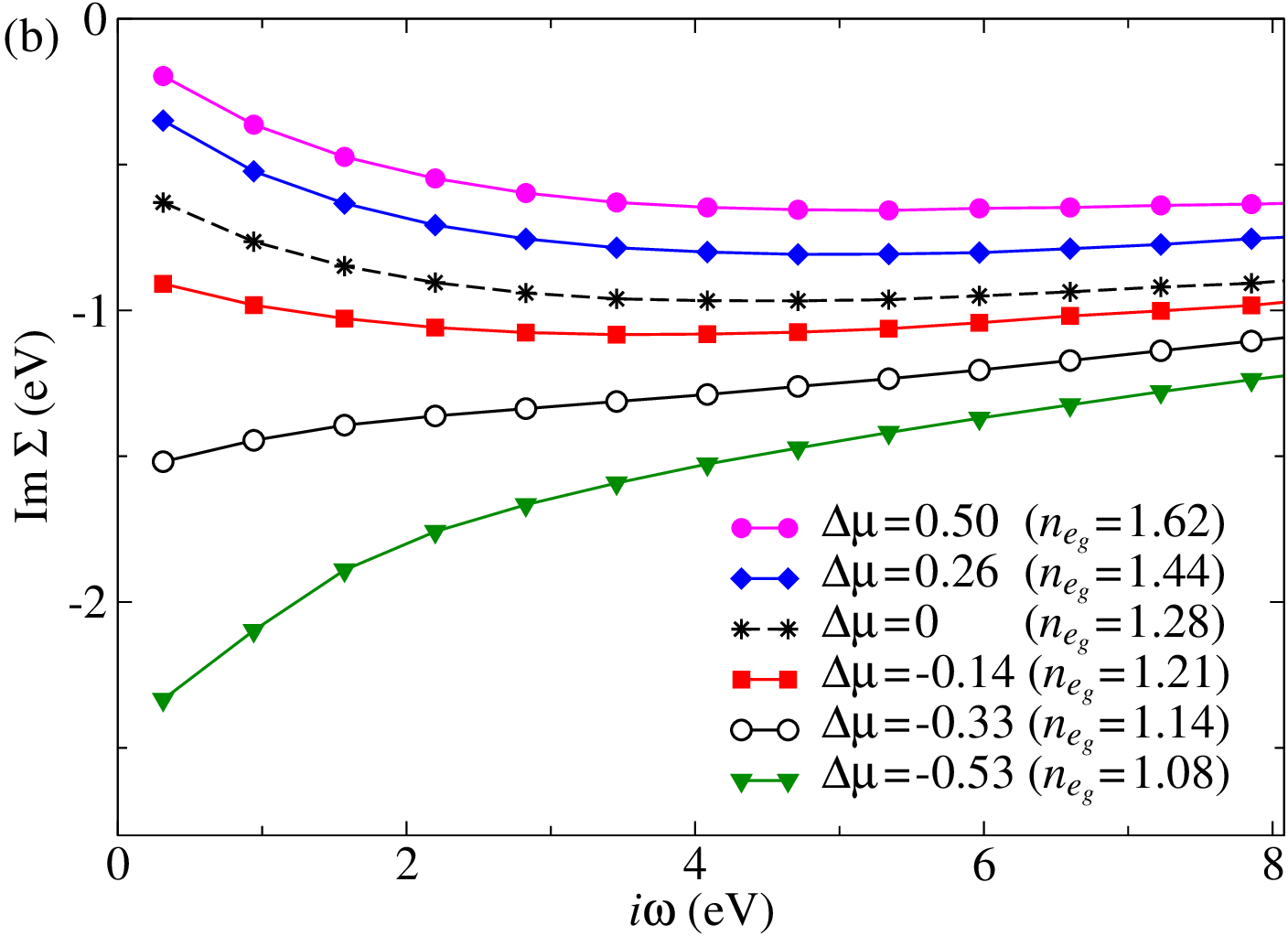}
\caption{(Color online) The same as in Fig. \ref{fig:sigma_bcc} for $\gamma$-iron. 
\label{fig:sigma_fcc}}
\end{figure}

To study the effect of filling on the self-energies of $\alpha$- and $\gamma$-iron, 
we also consider the results of DMFT calculations of the five-band model with the \textit{ab initio} dispersion, which yields density of states, presented in Fig. 4,
but a different 
concentration of $d$ 
electrons
achieved by a shift of the chemical potential $\Delta \mu$ with respect to its position in $\alpha$- or $\gamma$-iron.  As in Sect. II, we also assume a local form of the Coulomb interaction with the same parameters as in the previous study of Ref.~\cite{Katanin2016}. 
We find (see Figs. \ref{fig:sigma_bcc},\ref{fig:sigma_fcc}) that decreasing filling towards half filling always increases quasiparticle damping, yielding peculiarities of the local susceptibility, described above for the two-band model.
The most pronounced change with changing the filling corresponds to $e_g$ electrons in $\gamma$-iron (see Fig.~\ref{fig:sigma_fcc}b).
For the filling, corresponding to the pure $\gamma$-iron, the $e_g$ states are characterized by quasiparticle-like self-energy, with $|{\rm Im}\Sigma(i \omega)|$ decreasing with decreasing $\omega$ (see the self-energy for ${\Delta \mu=0}$ 
in Fig.~\ref{fig:sigma_fcc}b, cf. Ref.~\cite{OurGamma}). A reduction of 
number of $e_g$ electrons per orbital towards~1
(corresponding to half filling) monotonically increases
the absolute value of the self-energy ultimately yielding for ${n_{e_g}<1.18}$ 
above half filling non-quasiparticle like self-energy, despite the Fermi level shifting further from the peak. 

Because of stronger asymmetry and larger density of states at the Fermi level (which occur due to the proximity to the peak of the density of states for $e_g$ states), for comparable fillings per orbital $e_g$ states in $\alpha$ iron have stronger quasiparticle damping (or even show non-quasiparticle behavior), than the $t_{2g}$ states. The same applies to $\gamma$-iron close to half filling (see, for example, ${n_{e_g}=1.14}$ and ${n_{t_{2g}}=1.15}$ self-energies), but for larger fillings the situation there is more complex, possibly because of the substantial hybridization between $e_g$ and $t_{2g}$ states in this substance (see the Appendix).  

\begin{figure}[t]
\includegraphics[clip=true,width=0.43\textwidth]{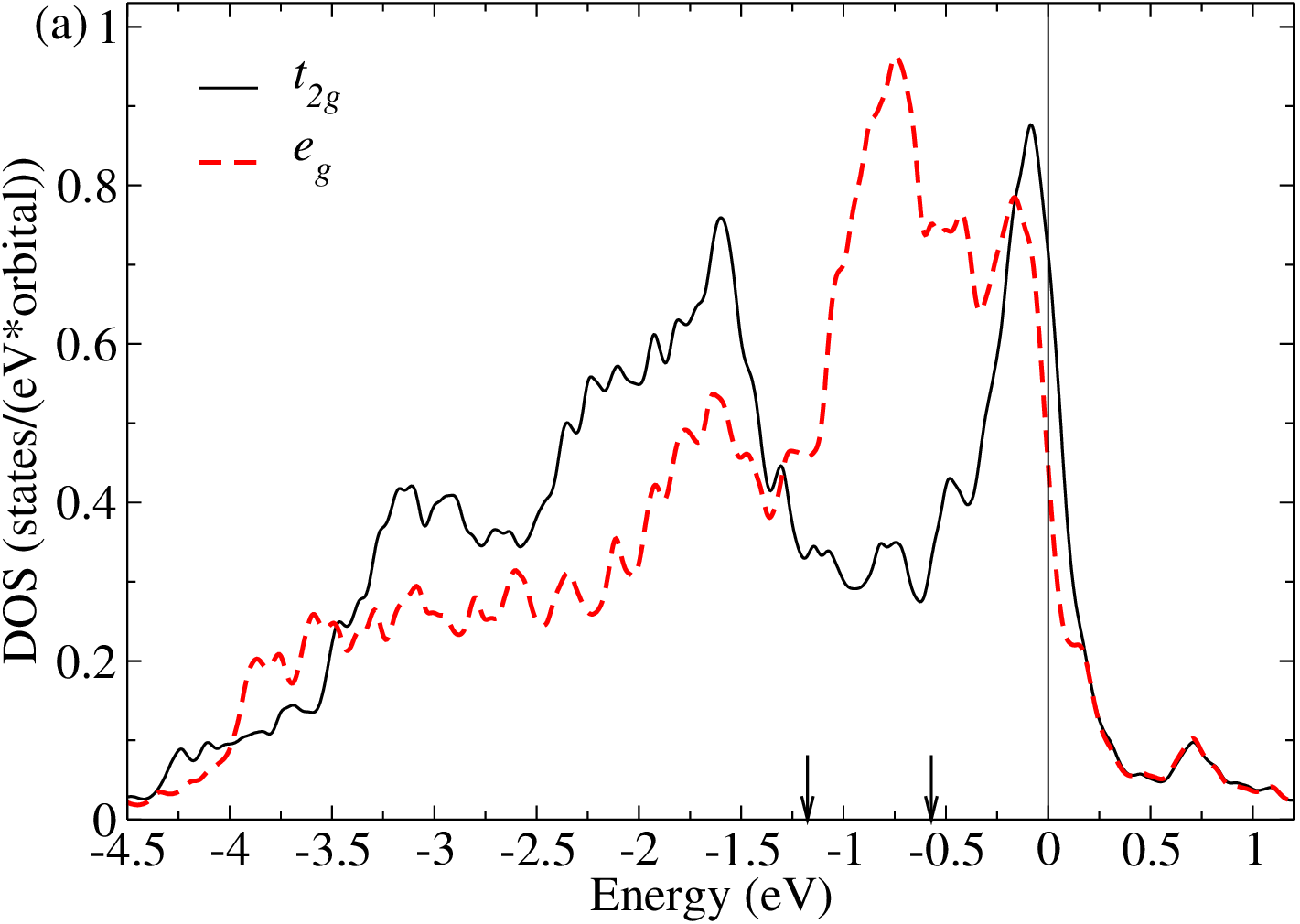}
\includegraphics[clip=true,width=0.435\textwidth]{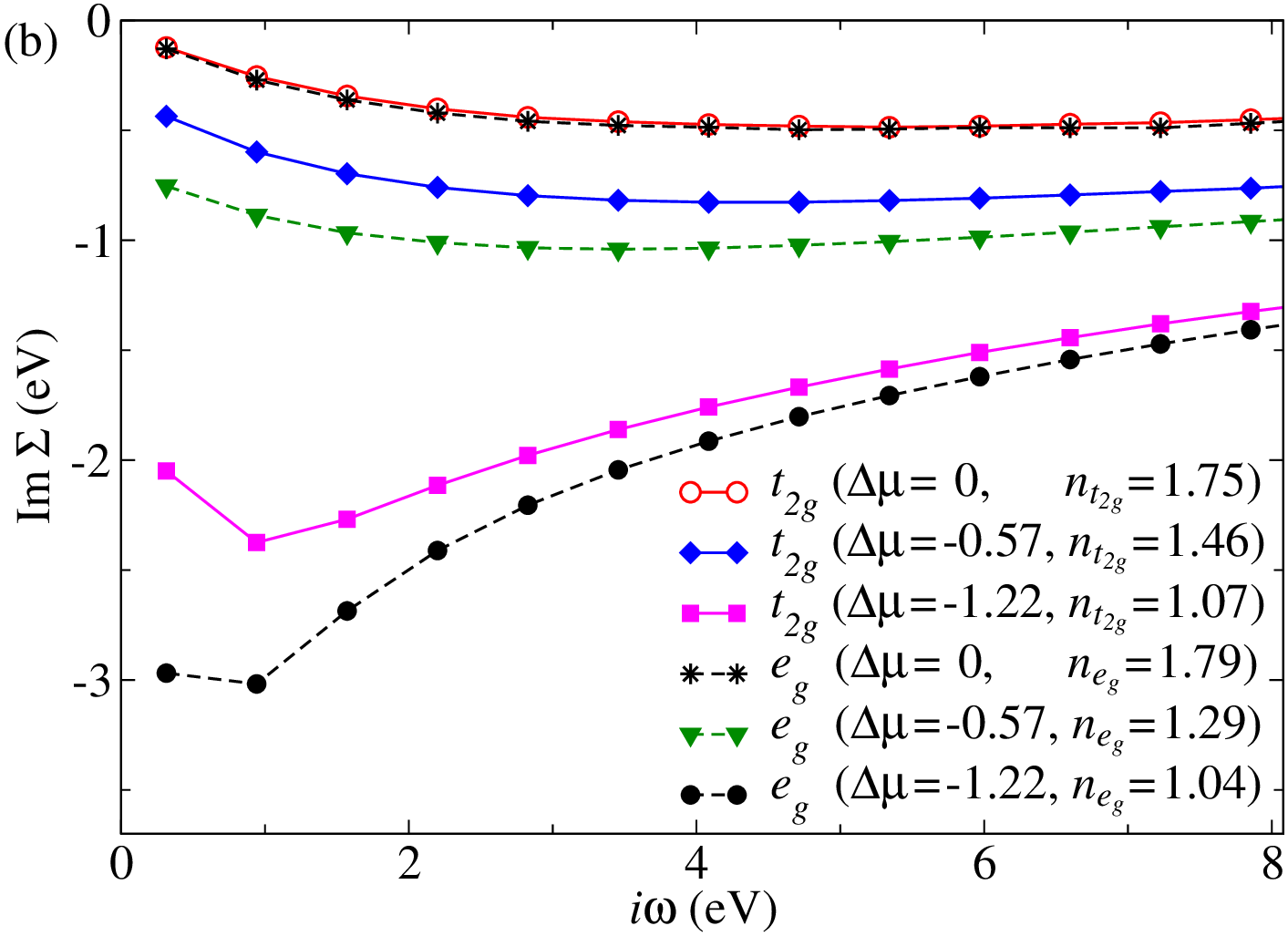}
\caption{(Color online) Partial densities of states (a)  and self-energies (b) of Ni for different fillings and $\beta=10$~eV$^{-1}$.
\label{fig:sigma_ni}}
\end{figure}

In nickel the partial density of $e_g$ states is smaller at the Fermi level, than the one for $t_{2g}$ states, but has stronger asymmetry 
(see Fig.~\ref{fig:sigma_ni}a and Table~I). For the \textit{ab initio} position of the Fermi level, the filling is very far from half filling (${n=8.54}$), and therefore the local moments are not fully formed in this substance; it is also characterized by large Kondo temperature ${T_\textrm{K} \sim 600}$~K \cite{Sangiovanni}. The dependence of the self-energy on the position of the chemical potential is determined mainly by the filling of the $e_g$ and $t_{2g}$ states (see Fig.~\ref{fig:sigma_ni}b), yielding stronger quasiparticle damping 
on approaching half filling.

Finally we mention some layered materials. In iron pnictides the partial DOS are asymmetric, but the local moments are partially formed mainly due to the proximity of some bands to half filling, at least in LaFeAsO mostly studied in that respect \cite{OurPnic,Toschi}.
For cuprates, the effective two-band model, 
derived to account for non-local correlations beyond DMFT \cite{SF_cuprates}, 
has an asymmetric density of states 
for the non-vanishing next-nearest hopping $t'$ 
and the Fermi level is shifted towards the maximum of the density of states, which may also enhance spin freezing near half filling. In this model, however, the bands are not degenerate, and therefore this case requires further investigations. Since the effective model corresponds to the non-local (plaquette) degrees of freedom of the original model, the spin freezing in this case does not necessarily correspond to the local moment formation, and may be accompanied by non-trivial behavior of the charge degrees of freedom (e.g., charge order, orbital currents, phase separation, etc.). 


\section{Conclusions}

In summary, we have investigated the effect of asymmetry of the density of states on Hund's metal behavior in multiband Hubbard models. We find, that the asymmetry and larger value of the density of states at the Fermi level enhance the Hund's metal behavior, i.e., yield stronger quasiparticle damping (or non-quasiparticle behavior) and stronger spin freezing, corresponding to the formation of local moments, because of the stronger correlated and larger number of electron or hole excitations. The inverse local susceptibility becomes linear in temperature with the Kondo scale decreasing with increasing the asymmetry. The above discussed behavior is observed sufficiently close to half filling. 
For realistic densities of states the two considered factors (asymmetry and value of the density of states at the Fermi level) are often interrelated, since the Fermi level is attracted to the peaks of the density of states in a broad range of fillings due to both, band structure and correlation effects.

In the present study we did not consider the effect of the non-local correlations, but they are not expected to qualitatively change the obtained results, except, possibly, for the narrow critical regions in the vicinity of the ordered phases, which require further investigation. The obtained features allow us, on one hand, to explain properties of known materials, but on the other hand, they allow one to predict the way to find new materials showing Hund's metal behavior and, therefore, unusual electronic and magnetic properties.




The work was supported by the Russian Science Foundation (project no. 14-22-00004).

\setcounter{table}{0}
\renewcommand\thetable{A\Roman{table}}
\appendix
\section{Hybridization between $e_g$ and $t_{2g}$ states in iron and nickel}
As a measure of the strength of the hybridization between $e_g$ and $t_{2g}$ states, we consider the kinetic energy contribution
\begin{equation}
E_{\text{kin}}^{MM^{\prime }}=\sum\limits_{\mathbf{k},m\in M,m^{\prime }\in
M^{\prime },\sigma }H_{\mathbf{k}}^{mm^{\prime }}\langle c_{\mathbf{k}m\sigma
}^{+}c_{\mathbf{k}m^{\prime }\sigma }\rangle 
\label{Eq:Ekinmm1}
\end{equation}
where $c_{\mathbf{k}m\sigma }^{+}$ ($c_{\mathbf{k}m\sigma }$) are the operators of the creation
(annihilation) of the electron with momentum $\mathbf{k}$ and spin $\sigma $ at the
orbital $m,$ $H_{\mathbf{k}}^{mm^{\prime }}$ is the \textit{ab initio} Hamiltonian in
momentum space (estimated with respect to the Fermi level) and $M,M^{\prime }$ denote states of different symmetries ($%
e_{g}$ and $t_{2g}$ for cubic lattice). The \textit{ab initio} calculations with the average in Eq. (\ref{Eq:Ekinmm1}) obtained with the Hamiltonian $H_{\mathbf{k}}^{mm^{\prime }}$ yield the
results, presented in Table AI. One can see that $|E_{\text{kin}%
}^{e_{g},t_{2g}}|\ll |E_{\text{kin}}^{e_{g},e_{g}}|,$ $|E_{\text{kin}%
}^{t_{2g},t_{2g}}|.$

\begin{table}[h]
\vspace{0.0cm}
\begin{ruledtabular}
\begin{tabular}{cccc}
& $\alpha $-Fe & $\gamma $-Fe & Ni \\ \hline
$E_{\text{kin}}^{e_{g},e_{g}}$   & $-$2.777 & $-$2.308 & $-$4.099 \\ 
$E_{\text{kin}}^{t_{2g},t_{2g}}$ & $-$5.045 & $-$4.223 & $-$6.673 \\ 
$E_{\text{kin}}^{e_{g},t_{2g}}$  & $-$0.118 & $-$0.550 & $-$0.243 \\ 
\end{tabular}%
\end{ruledtabular}
\caption{Kinetic energy of states of different symmetry in iron and nickel in eV}
\end{table}

\vspace{-0.5cm}


\begin{thebibliography}{9}    

\bibitem{HM} Z. P. Yin, K. Haule, and G. Kotliar, Nature Mater. {\bf 10}, 932 (2011).

\bibitem{Medici} L. de' Medici, J. Mravlje, A. Georges, Phys. Rev. Lett. {\bf 107}, 256401 (2011).

\bibitem{Medici1} L. de' Medici, Phys. Rev. B {\bf 83}, 205112 (2011).

\bibitem{Pruschke} T. Pruschke and R. Bulla, Eur. Phys. J. B {\bf 44}, 217 (2005).

\bibitem{SF} P. Werner, E. Gull, M. Troyer, and A. J. Millis, Phys. Rev. Lett. {\bf 101}, 166405 (2008).

\bibitem{Licht2001} A. I. Lichtenstein, M. I. Katsnelson, and G. Kotliar,
Phys. Rev. Lett. {\bf 87}, 067205 (2001).

\bibitem{OurAlpha0} A. A. Katanin, A. I. Poteryaev, A. V. Efremov, A. O. Shorikov, S. L. Skornyakov, M. A. Korotin, V. I. Anisimov, Phys. Rev. B \textbf{81}, 045117 (2010).

\bibitem{OurAlpha}P. A. Igoshev, A. V. Efremov, A. A. Katanin, Phys. Rev. B \textbf{91}, 195123 (2015).

\bibitem{Belozerov2013}
  A. S. Belozerov, I. Leonov and V. I. Anisimov, Phys. Rev. B \textbf{87}, 125138 (2013).
  
\bibitem{Pourovskii} L. V. Pourovskii, T. Miyake, S. I. Simak, A. V. Ruban, L. Dubrovinsky, I. A. Abrikosov, Phys. Rev. B \textbf{87}, 115130 (2013).

\bibitem{Sangiovanni} A. Hausoel, M. Karolak, E. Sasioglu, A. Lichtenstein, K. Held, A. Katanin, A. Toschi,
and G. Sangiovanni, Nature Comm. {\bf 8}, 16062 (2017).

\bibitem{OurGamma}P. A. Igoshev, A. V. Efremov, A. I. Poteryaev, A. A. Katanin, V. I. Anisimov, Phys. Rev. B \textbf{88}, 155120 (2013).

\bibitem{Toschi} P. Hansmann, R. Arita, A. Toschi, S. Sakai, G. Sangiovanni, and K. Held, Phys. Rev. Lett. {\bf 104}, 197002 (2010); A. Toschi, R. Arita, P. Hansmann, G. Sangiovanni, K. Held, Phys. Rev. B {\bf 86}, 064411 (2012).

\bibitem{OurPnic} S. L. Skornyakov, A. A. Katanin, and V. I. Anisimov,
Phys. Rev. Lett. {\bf 106}, 047007 (2011).  

\bibitem{Anisimov} V. I. Anisimov, I. A. Nekrasov, D. E. Kondakov, T. M. Rice, and M. Sigrist, Eur. Phys. J. B {\bf 25}, 191 (2002).

\bibitem{OSMT1} A. Koga, N. Kawakami, T. M. Rice, and M. Sigrist, Phys. Rev. Lett. {\bf 92}, 216402 (2004).
\bibitem{OSMT2}  S. Biermann, L. de Medici, and A. Georges, Phys. Rev. Lett. {\bf 95}, 206401 (2005).
\bibitem{OSMT3}  A. Liebsch, Phys. Rev. Lett. {\bf 95}, 116402 (2005).
\bibitem{OSMT4}  C. Knecht, N. Bl\"umer, and P. G. J. van Dongen, Phys. Rev. B {\bf 72}, 081103(R) (2005).
\bibitem{OSMT5}  R. Arita and K. Held, Phys. Rev. B {\bf 72}, 201102(R) (2005).
\bibitem{OSMT6}  T. A. Costi and A. Liebsch, Phys. Rev. Lett. {\bf 99}, 236404 (2007).
\bibitem{OSMT7}  L. de Medici, S. R. Hassan, M. Capone, and X. Dai, Phys. Rev. Lett. {\bf 102}, 126401 (2009).

\bibitem{Pnic1} A.O. Shorikov, M.A. Korotin, S.V. Streltsov, S.L. Skornyakov, D.M. Korotin, V.I. Anisimov, Journ. Exp. Theor. Phys. {\bf 108}, 121 (2009).

\bibitem{Pnic2} V. I. Anisimov, Dm. M. Korotin, M. A. Korotin, A. V. Kozhevnikov, J. Kuneš, A. O. Shorikov, S. L. Skornyakov, S. V. Streltsov, 	J. Phys.: Condens. Matter {\bf 21}, 075602 (2009).

\bibitem{Pnic3} S.L. Skornyakov, A.V. Efremov, N.A. Skorikov, M.A. Korotin, Yu.A. Izyumov, V.I. Anisimov, A.V. Kozhevnikov, D. Vollhardt, Phys. Rev. B {\bf 80}, 092501 (2009).

\bibitem{Pnic4} S.L. Skornyakov, N.A. Skorikov, A.V. Lukoyanov, A.O. Shorikov, V.I. Anisimov, Phys. Rev. B {\bf 81}, 174522  (2010).

\bibitem{Pnic5} S.L. Skornyakov, V.I. Anisimov, D. Vollhardt, JETP Lett. {\bf 100}, 120 (2014).

\bibitem{CT-QMC} A. N. Rubtsov, V. V. Savkin, and A. I. Lichtenstein, Phys. Rev. B \textbf{72}, 035122 (2005);
P. Werner, A. Comanac, L. de Medici, M. Troyer, and A. J. Millis, Phys. Rev. Lett. \textbf{97}, 076405 (2006).

\bibitem{Wahle} J. Wahle, N. Bl\"umer, J. Schlipf, K. Held, and D. Vollhardt, Phys. Rev. B \textbf{58}, 12749 (1998).

\bibitem{Medici1a} A. Georges, L. de' Medici, and J. Mravlje, Annual Rev. Cond. Matter Phys. {\bf 4}, 137 (2013).

\bibitem{Medici2} L. de' Medici and M. Capone in \textit{The Iron Pnictide Superconductors}, edited by F. Mancini and R. Citro, Springer Series in Solid-State Sciences (2017), Vol. 186, p. 115.

\bibitem{Medici3} L. de' Medici, in \textit{The Physics of Correlated Insulators, Metals, and Superconductors Modeling and Simulation}, edited by E. Pavarini, E. Koch, R. Scalettar, and R. Martin, Vol. 7 (Forschungszentrum J\"ulich, Germany, 2017).




\bibitem{Katanin2016}A. A. Katanin, A. S. Belozerov, and V. I. Anisimov, Phys. Rev. B \textbf{94}, 161117(R) (2016).

\bibitem{Wilson} K. G. Wilson,
Rev. Mod. Phys. {\bf 47}, 773 (1975).

\bibitem{Note} Note that we use the mapping of the lattice problem to local one and study the local (impurity) susceptibility to characterize the formation of local moments in the lattice problem. We do not consider here the question of finding the coherence temperature of the electrons on the lattice, at which the Fermi liquid behavior sets in, and which may be different from the Kondo temperature of the individual local moments. 

\bibitem{Alpha_supercell} A. S. Belozerov, A. A. Katanin, V. I. Anisimov, Phys. Rev. B {\bf 96}, 075108 (2017).

\bibitem{SF_cuprates} P. Werner, S. Hoshino, and H. Shinaoka,
Phys. Rev. B {\bf 94}, 245134 (2016).
\end{thebibliography}
\end{document}